%%%%%%%%%%%%%%%%%%%%%%%%%%%%%%%%%%%%%%%%%%%%%%%%%%%%%%%%%%%%%%%%%%%%%%%%%%%
% Title:   Bispectral Operators of Rank $1$ and   
%          Dual Isomonodromic Deformations 
% Authors: J. Harnad 
% Date:    December 16, 1996
%=============================================================
% Type:    Preprint CRM-2443 (1996) / Conference Proceedings
%          (To appear in: CRM 25th Anniverary Volume (1996)          
%=============================================================
% Compiler:  AMSTeX version 2.1 or later
% ===========================================================
%%%%%%%%%%%%%%%%%%%%%%%  Formatting Specifications  %%%%%%%%%%%%%%%%%%%%%
\documentstyle{amsppt}
%\NoPageNumbers
\TagsOnRight
% The following 3 lines "disable" the printing of the AMSTeX logo.
% Please be sure that proper acknowledgment is made upon publication.
\catcode`\@=11
\def\logo@{}
\catcode`\@=13
\parindent=8 mm
\magnification 1200
\hsize = 6.25 true in
\vsize = 8.5 true in
\hoffset = .25 true in
\parskip=\medskipamount
\baselineskip=14pt
\phantom{0}
\vskip -.5 true in 
\noindent 
CRM-2443 (1996)
\hfill solv-int/9612012 
\break 
\bigskip 
%%%%%%%%%%%%%%%%%%%%%%%%%%%% Definitions %%%%%%%%%%%%%%%%%%%%%%%%%%%%%%%%
\def \oc {\overset {\otimes}\to{,}}
\def \di{\partial}

\def \smaller {\eightpoint}
\def \wt {\widetilde}

\def \mt {\mapsto}
\def \ra {\rightarrow}

\def \lra {\longrightarrow}
\def \lmt {\longmapsto}
\def \a {\alpha}
\def \b {\beta}
\def \d {\delta}

\def \g {\gamma}
\def \G {\Gamma}

\def \l {\lambda}
\def \L {\Lambda}

\def \th {\theta}
\def \t {\tau}
\def \o {\omega}

\def \ss {\subset}

\def \GGG {{\frak G}}

\def \Ggl {\frak{gl}}
\def \GGL {\frak{Gl}}

\def \CB {\bold{C}}

\def \IB {\bold{I}}

\def \cB {\bold{c}}

\def \tB {\bold{t}}
\def \uB {\bold{u}}
\def \vB {\bold{v}}

\def \AA {\Cal A}

\def \DD {\Cal D}

\def \HH {\Cal H}

\def \MM {\Cal M}
\def \NN {\Cal N}

\def \WW {\Cal W}

\def \di {\partial}

\def \ln{\text{ln}}
\def \tr{\text{tr}}

\def \diag{\text{diag }}
\def \End{\text{End}}

\hyphenation{non-au-to-nomous equations bundle}
\leftheadtext{J\. Harnad}
\rightheadtext{Bispectral Operators and Isomonodromy}
%%%%%%%%%%%%%%%%%%%%%%%%%%%%%%%%%% Title %%%%%%%%%%%%%%%%%%%%%%%%%%%%%%%%%
\topmatter
\title
Bispectral Operators of Rank $1$ and Dual Isomonodromic Deformations${}^\dag$
\endtitle
\thanks \noindent ${}^\dag$Research supported in part by the Natural Sciences and
Engineering Research Council of Canada and the Fonds FCAR du Qu\'ebec.
To appear in the C.R.M. 25th Anniversary Volume (1996/97). 
\endthanks
\author
J\. Harnad
\endauthor
\affil
\smaller{
Department of Mathematics and Statistics, Concordia
University \\
7141 Sherbrooke W., Montr\'eal, Canada H4B 1R6, and \\
Centre de recherches math\'ematiques, Universit\'e de Montr\'eal \\
 C\. P\. 6128, Succ. centre--ville, Montr\'eal, Canada H3C 3J7 \\
e-mail: {\it harnad\@alcor.concordia.ca}  or  {\it harnad\@crm.umontreal.ca}}
\endaffil
%%%%%%%%%%%%%%%%%%%%%%%%%%%%%%%%%  Abstract  %%%%%%%%%%%%%%%%%%%%%%%%%%%%%
\abstract
A comparison is made between bispectral operator pairs and dual pairs of 
isomonodromic deformation equations.  Through examples, it is shown how 
operators belonging to rank one bispectral algebras may be viewed
equivalently as defining $1$--parameter families of rational first order
differential operators with matricial coefficients on the Riemann sphere,
whose monodromy is trivial. By interchanging the r\^oles of the two variables
entering in the bispectral pair, a second 1-parameter family of operators
with trivial monodromy is obtained, which may be viewed as the dual
isomonodromic deformation system. 
 \endabstract
\endtopmatter
\document 
\baselineskip=14pt 
%%%%%%%%%%%%%%%%%%%%%%%%%%%%%%% Section 1  %%%%%%%%%%%%%%%%%%%%%%%%%%%%%%%%
\subheading{1. Bispectral Operators}
\nopagebreak \smallskip
\line{1.1 \quad{\it Bispectral Pairs.}\hfil}
 \nopagebreak

  Consider pairs of differential operators $L(x, \di_x), \L(z, \di_z)$ in 
two variables $x$ and $z$, for which there exists a function $\psi(x,z)$
that is simultaneously a parametric family of eigenfunctions of $L$ and $\L$.
$$
\align
L\psi(x,z) &= f(z) \psi(x,z) \tag{1.1a}\\
\L \psi(x,z) &= \phi(x)\psi(x,z),    \tag{1.1b}
\endalign 
$$
where the eigenvalues $f(z)$ and $\phi(x)$ are nonconstant functions of the
variables $z$ and $x$. Such pairs of operators were named
{\it bispectral} and studied by Duistermaat and Gr\"unbaum in [DG]. Choosing 
a normalization in which both $L$ and $\L$ have unit leading coefficients and 
constant next to leading coeffients (which is always possible, up to
reparametrization), they were able to draw a number of conclusions about
these operators. In particular,   the coefficents of $L$ and $\L$ are
rational in the variables $x$ and $z$, respectively, and the functions $f(x)$
and $\phi(z)$ are polynomials.  Furthermore, for the case of second order
operators: $$
L = {d^2 \over dx^2} +u(x)  \tag{1.2}
$$
[DG] were able to determine all possible $u(x)$; namely (up to
translations in  $x$ or addition of a constant),
$$
u(x) = x \quad (Airy)  \qquad \text{or} \qquad
u(x)       = {c \over x^2} \quad (Bessel)   \tag{1.3}
$$
or anything that can be obtained from the two cases
$$
u(x) = 0 , \qquad u(x) = -{1 \over 4x^2}   \tag{1.4}
$$
through the application of rational Darboux transformations.
\medskip
\line{1.2 \quad{\it Bispectral Algebras of Rank $1$.\hfil}} 
\nopagebreak

   Wilson [W1, W2] considered the case when $L$ is embedded in a commutative 
algebra $\AA$ of differential operators, sharing the same family of
eigenfunctions  $\psi(x,z)$, such that the orders of the elements of $\AA$
are relatively prime (a rank $1$ algebra). Such algebras can be characterized
by the associated spectral data, consisting of an algebraic curve, the
spectral curve of $\AA$, denoted spec$(\AA)$ and, in general, a line bundle
whose fibres are the joint eigenvectors. This data may be viewed as
determining a point $W\in \text{Gr}$ in the Hilbert space Grassmannian of
Sato [SS] and Segal--Wilson [SW], such that the bispectral wave function
$\psi(x,z)$ is the corresponding Baker--Akhiezer function $\psi_W(\tB, z)$ at
the point $\tB :=(t_1,t_2,\dots)$ with $(t_1=x, t_i=0,\ i \ge 2)$.   We
recall that the Baker-Akhiezer function is the unique function of the form   
$$
\psi_W(\tB,z) 
= \g(\tB) \left( 1 + \sum_{i=1}^{\infty} {b_i(\tB) \over z^i}\right),
\tag{1.10} 
$$  
where $\g(\tB):= e^{\sum_{j=1}^\infty t_j z^j}$, taking values in
the subspace  $W$ of the Hilbert space $\HH:= L^2(S^1,\CB)$ of square
integrable functions on  the  unit circle in the complex $z$--plane. Here
$\tB=(t_1, t_2, \dots)$ is an infinite component vector and $\g(\tB)$ may be
viewed as an element of the infinite abelian group $\G_+$ consisting of
elements of $L^2(S^1, \CB)$ admitting a holomorphic continuation to the
interior of $S^1$ in the complex $z$--plane and taking value $1$ at the
origin. The Baker--Akhiezer function may, in turn, be expressed in terms of
the Sato tau function associated to $W$, by the formula [DJKM] $$ 
\psi_W(\tB,z) = \g(\tB) {\tau_W(\tB - \bold{[z]})
\over \tau_W(\tB)},  \tag{1.11} 
$$
where the components of $\bold{[z]}= (z_1, z_2, \dots)$ are 
$$
z_k := {1 \over k z^k}.   \tag{1.12}
$$
Geometrically, $\t_W$ is understood as the determinant, defined up to 
normalization,  of the projection $P_+:\g W \ra \HH_+$, where $\g W$ is the
image under $\g\in \G_+$ of the subspace $W\ss \HH$ and $\HH_+ \ss \HH$ is
the subspace consisting of elements admitting a holomorphic extension to the
enterior of $S^1$.

 Wilson [W1] gave the following equivalent characterizations of such $W$'s
corresponding to bispectral algebras of rank $1$.
 \proclaim{Theorem (Wilson [W1])} The algebra $\AA$ is bispectral and of rank
$1$ if  and only if the following equivalent conditions hold:
\item {(i)} $\text{spec}(\AA)$ is a rational algebraic curve with only cusp
singularities.
\item{(ii)} Within the proper normalization, the Baker--Akhiezer function
$$
\psi(x,z) = \psi_W(x=t_1, t_j=0,j \ge 2, z)
= e^{xz} \left( 1 + \sum_{i=1}^{\infty} {b_i(x) \over z^i}\right)  \tag{1.13}
$$ 
belongs to a plane $W$ belonging to a subvariety $\text{Gr}^{\text{ad}}$ of
the {\it rational} Segal--Wilson Grassmannian $\text{Gr}^{\text{rat}}$,
called the  {\it adelic} Grassmannian (defined below). 
\endproclaim
 The adelic Grassmannian $\text{Gr}^{\text{ad}}$ is shown in [W1] to consist
of the  $W$--planes in the Hilbert space $\HH$ of the form
$$
W= {1\over q(z)} W_+^{\cB}  \tag{1.14}
$$
where $q(z)$ is a polynomial of suitable degree, rendering the projection 
$P_+:W\ra \HH_+$ to the subspace $\HH_+\ss \HH$ a Fredholm operator of index
$0$, and the subspace $W_+^{\cB}\ss \HH_+$ is the intersection of the kernels
of a finite number of linear forms \{$c^{\mu}_i \in \HH_+^*, \}$, each of
which has support at one point $\l_i$ in the complex $z$--plane. That is,
each $c^{\mu}_i$ determines a {\it condition} of the form   
$$
c^{\mu}_i (g) = \sum_{a=1}^{m^\mu_i}c^{\mu}_{ia}g^{(a)}(\l_i)=0,  \tag{1.15}
$$
where $g^{(a)}$ denotes the $a$--th derivative of $g\in \HH_+$ and 
$\{c^\mu_{ia}\}$ is a finite set of coefficients. The $\l_i$'s are just the
roots of the polynomial $q(z)$, with multiplicities coinciding with the
number of conditions localized at that root.

   Wilson also showed that the property of bispectrality could be viewed as a
consequence of the existence of a ``bispectral involution''
$$
\aligned
b: \text{Gr}^{\text{ad}} &\ra \text{Gr}^{\text{ad}} \\
b: W &\mt W^\prime  
\endaligned  \tag{1.16}
$$
such that 
$$
\psi_{W^\prime}(x,z) = \psi_W(z,x)  \tag{1.17}
$$

   The larger, {\it rational} Grassmannian $\text{Gr}^{\text{rat}}$ contains
all rational and soliton solutions to the KP hierarchy. It was shown by
Krichever [Kr], following earlier work of Airault, Mckean and Moser [AMM] on
rational solutions to the KdV equation, that rational solutions of the KP
equation which tend to zero as $x\ra \infty$ can be expressed as  
$$
u(x,y,t) = -\sum_{i=1}^n {2 \over (x- x_i(y,t))^2},   \tag{1.18}
$$
where the location of the poles $x_i(y,t)$ is determined by the fact that
they  satisfy the equations of the first two commuting flows of the
Calogero-Moser system [M]. It follows that these are associated to bispectral
tau functions $\tau_W$. In [W2], Wilson showed that, by suitably completing
the complexified Calogero-Moser phase space  so as to allow for particle
collisions, this extends to a complete parametrization of the adelic
Grassmannian $\text{Gr}^{\text{ad}}$, and hence the set of rank $1$
bispectral algebras. More precisely, let $C_n$ denote the completed
$n$-particle rational Calogero--Moser phase space, defined as the space of
pairs $(X,Z)$ of complex $n \times n$ matrices whose commutator $[Z,X]$ is a
rank $1$ perturbation of the identity matrix $\IB$, quotiented by the natural
action of the general linear group $\GGL(n, \CB)$ on such pairs; $$
g: (X,Z) \mt (gXg^{-1}, gZg^{-1}), \quad g \in \GGL(n, \CB).   \tag{1.19}
$$
Define a map from the union of the spaces $C_n$ to the adelic Grassmannian
$$
\aligned
\WW: \cup_n C_n &\ra \text{Gr}^{\text{ad}} \\
\WW: [(X,Z)] & \mt W(X,Z),
\endaligned   \tag{1.20}
$$
where $W(X,Z)\in \text{Gr}^{\text{ad}}$ is determined by the following
formula  for the corresponding Baker function at $\tB=(x,0,\dots)$.
$$
\psi_{W(X,Z)}(x,z) = e^{xz} \det\left(\IB - (X + x\IB)^{-1} 
(Z+ z\IB)^{-1}\right).
\tag{1.21}
$$
(Note that $\psi_W(x,z)$ and its derivatives at $x=0$ span $W$.)
Equivalently,  the corresponding tau function may be expressed 
$$
\tau_{W(X,Z)}= \det(X + x\IB + \sum_{j=2}^\infty j t_j (- Z)^{j-1}). 
\tag{1.22} 
$$
  Wilson showed in [W2] that this identification gives an isomorphism of
affine, nonsingular, irreducible algebraic varieties. It is clear from
(1.16), (1.21)  that under this identification, the bispectral involution is
equivalent to the symplectic involution   
$$
\aligned
\tilde{b}: C_n &\ra C_n \\
\tilde{b}: (X,Z) &\mt (Z^t, X^t)
\endaligned  \tag{1.23}
$$
on the completed Caloger--Moser phase space. (This was noted for the case of
the original real Calogero--Moser phase by Kasman [K], with $\tilde{b}$
viewed as Moser's linearizing involution [M].)  

  This therefore provides a complete determination of all wave functions
$\psi_W$ corresponding to rank $1$ bispectral algebras. The following are two
of the simplest examples of these. (In each case, the algebra $\AA$ is
identified  with the space of polynomials which, under multiplication, leave
$W(X,Z)$ invariant. Only the simplest representative elements $L$ and $\L$,
are given.)

\noindent {\it Example (i)}
$$
\align   n=1, \quad X&=(\a), \quad Z= (0)   \tag{1.24a}\\
\psi_{W(X,Z)}(x,z) &= e^{xz} \left( 1 - {1\over (x+ \a)z}\right)
\tag{1.24b}\\ 
L&= {d^2\over dx^2} - {2\over (x+\a)^2}, \quad f(z)=z^2  
\tag{1.24c}\\ \L&={d^2\over dz^2}+2\a {d\over dz} - {2\over z^2}, \quad
\phi(x)=x^2+2\a x \tag{1.24d} \endalign
$$
\newline {\it Example (ii)}
$$ 
\align n=2, \quad X &=\pmatrix 
0 & 0\\ 1 & 0
\endpmatrix,
 \quad Z= \pmatrix 
0 & -1 \\ 0 & 0 
\endpmatrix  \tag{1.25a}\\
\psi_{W(X,Z)}(x,z) &= e^{xz} \left(1 - {2\over x z} + {2\over x^2 z^2}\right)
\tag{1.25b}
\\
L&= {d^3\over dx^3} - {6\over x^2}{d\over dx}+{12\over x^2}, \quad f(z)=z^3 
\tag{1.25c}
\\
\L&={d^3\over dz^3} - {6\over z^2}{d\over dz}+{12\over z^2},  \quad
\phi(x)=x^3 \tag{1.25d}
\endalign
$$

 \bigskip
\subheading{2. Isomonodromic Deformations} 
\nopagebreak\smallskip
\line{2.1 \quad{\it Rational Covariant Derivatives Operators.}\hfil}
 \nopagebreak

Consider a parametric family of matrix first order differential operators
$$
\DD_\l(\t) := {\di \over \di \l} - B(\l, \t) -\sum_{i=1}^n\sum_{a=1}^{m_i}
{N_{ia}(\t)\over (\l-\a_i(\t))^a}, \tag{2.1}
$$
where $B(\l,\t)$ is an $r \times r$ matrix valued polynomial in $\l$ that may 
depend also on a deformation parameter $\t$, 
$\{N_{ia}\}_{1\le i \le n, 1\le a\le m_i}$ is a set of $r \times r$ matrices,
also dependent on $\t$, and  $\{\a_i(\t)\in \bold{C}\}_{1\le i \le n}$ are
the location of the finite poles, also possibly $\t$ dependent. This may be
viewed as a $\t$-parametric family of rational covariant derivative operators
defined on a rank $r$ vector bundle over the Riemann sphere, punctured at the
poles (including possibly $\infty$).  The generalized monodromy data
associated to the operator $\DD_\l$ consists of the local monodromy matrices
about each of the poles, together with the Stokes matrices and connection
matrices [JMU]. Assuming local differentiability in the deformation parameter
$\t$, it is a basic fact that this monodromy data is invariant under
variations in $\t$ provided there exists a second differential operator 
$$
\DD_\t := {\di \over \di \t} -E(\l, \t),  \tag{2.2}
$$
where $E(\l,\t)$ is also a $\t$--parametric family of $r \times r$ matrices, 
rational in $\l$, such the commutativity condition
$$
[\DD_\t, \DD_\l] =0  \tag{2.3}
$$
is satisfied. This means that the data $\{B(\l,\t), \a_i(\t), N_{ia}(\t)\}$
determining $\DD_\l(\t)$ satisfy a set of first order ODE's in the 
deformation parameter $\t$. It also implies that, locally at least, an
invertible matrix valued function $\Psi(\l, t)$ exists, uniquely determined
by any given set of initial values, simultaneously satisfying 
$$
\align
\DD_\l\Psi&=0  \tag{2.4a}\\
\DD_\t \Psi &= 0. \tag{2.4b}
\endalign
$$

   The simplest case, studied since the beginning of this century, is when 
the operator $\DD_\l$ is Fuchsian; i.e., it has only regular singular points.
This means that all the finite poles are simple ($ m_i=1, \ \forall i$) and
$B(\l,\t)\equiv 0$,   so $\DD_\l$ is of the form
$$
\DD_\l = {\di \over \di \l} -\sum_{i=1}^n{N_i \over \l -\a_i}. \tag{2.5}
$$
For this case, it has long been known that the only way the residue matrices
$\{N_i\}_{i=1,\dots n}$ can vary is through their dependence on the pole 
locations  $\{\a_i\}_{i=1,\dots n}$, and they are constrained to satisfy an
overdetermined system of PDE's.  Choosing the $\a_i$'s as the deformation
parameters, the associated infinitesimal isomonodromic deformation operators
are      $$ 
 \DD_{\a_i} := {\di \over \di \a_i} +{N_i \over  \l -\a_i}.  \tag{2.6} 
$$
The corresponding commutativity conditions
$$
[\DD_\l, \DD_{\a_i}] =0,  \quad [\DD_{\a_i}, \DD_{\a_j}] = 0, \quad 1\le, i,j 
\le n 
\tag{2.7} 
$$
are mutually compatible (i.e. integrable in the Frobenius sense), and
equivalent to the following  system of equations, known as the Schlesinger
equations $$
\align
{\di N_i \over \di \a_j} &= {[N_i, N_j]\over \a_i - \a_j}. \quad i\neq j 
\tag{2.8a}\\
{\di N_i\over \di \a_i} & = -\sum_{j=1 \atop j\neq i}^n {[N_i, N_j]
\over \a_i -\a_j}. \tag{2.8b}
\endalign
$$
(In particular, the simplest Fuchsian case involving nontrivial isomonodromic
deformations is when $n=3$ and $r=2$, for which these reduce to the equations of 
the sixth Painlev\'e transcendent $P_{VI}$.) 

  The corresponding determination of all equations of the type (2.3)
generating isomonodromic deformations of operators of the general form (2.1)
was made in  [JMU]. For the simplest nonfuchsian case, where the finite poles
remain first order, but an irregular singularity of Poincar\'e index $1$ is
added at infinity by allowing $B$ to be a nonzero diagonal matrix
$B(\t)=\diag(\b_1(\t), \dots \b_r(\t))$ with distinct eigenvalues that are
independent of $\l$, we have  $$
\DD_\l(\t)= {\di \over \di \l} - \NN(\l, \t)  \tag{2.9}
$$
where
$$
\NN(\l,\t) := B(\t) + \sum_{i=1}^n {N_i(\t) \over \l - \a_i}.  \tag{2.10}
$$
In this case, as shown in [JMMS], besides the original $\a_i$'s, the possible 
deformations parameters include the eigenvalues $\{\b_a\}_{1\le a \le r}$ of
the  matrix $B$. The corresponding infinitesimal isomonodromic deformation
operators  are  $$
\align
\DD_{\a_i}&:= {\di \over \di \a_i} +{N_i\over \l -\a_i}, \quad i=1, \dots, n 
\tag{2.11a} \\
\DD_{\b_a}&:= {\di\over \di \b_a} -\l E_a -\sum_{b=1 \atop b\neq a}^r  
{E_a N_\infty E_b + E_b N_\infty E_a \over \b_a -\b_b}, \quad a=1, \dots r,
\tag{2.11b} \endalign
$$
where $E_a$ is the elementary matrix with  $1$ in the $aa$ position and $0$'s
elsewhere, and
$$
N_\infty :=\sum_{i=1}^n N_i. \tag{2.12}
$$
Again the mutual commutativity of the operators 
$\{\DD_{\a_i}, \DD_{\b_a}, \DD_\l\}_{i=1,\dots n, a=1,\dots r}$,
$$
[\DD_{\a_i},\ \DD_{\a_j}]=[\DD_{\a_i}, \DD_{\b_a}] = [\DD_{\b_a},\
\DD_{\b_b}] = [\DD_\l, \DD_{\a_i}] = [\DD_\l, \DD_{\b_a}] = 0, \tag{2.13}
$$
gives a Frobenius integrable system for the residue matrices $N_i$, locally 
determining their dependence on the parameters $\{\a_i, \b_a\}$. 
\medskip
\line{2.2 \quad$R$--{\it Matrix theory and Duality.}\hfil}
 \nopagebreak
The above isomonodromic deformation equations may be viewed as nonautonomous
Hamiltonian equations ([JM], [H]) on the space $(\Ggl(r, \CB))^n$ of
$n$-tuples  of $r\times r$ matrices $\{N_1, \dots N_n\}$, with respect to the
Lie Poisson bracket structure  
$$
\{(N_i)_{ab},\ (N_j)_{cd}\}=\d_{ij}\left[\d_{ab} N_{cb} -\d_{bc}N_{da}\right].
\tag{2.14}
 $$
Equivalently, the rational matrix valued functions $\NN(\l)$, for 
different values of the complex parameters $\l$ and $\mu$, satisfy
$$
\{\NN(\l)\oc \NN(\mu)\}=\left[r(\l-\mu), \ \NN(\l) \otimes \IB + \IB\otimes
\NN(\mu)\right], \tag{2.15}
$$
where $r(\l-\mu)$ is the rational $R$--matrix, viewed as an element of
$\End(\CB^n \otimes \CB^n)$, defined by
$$
r(\l):= {P_{12}\over \l},  \tag{2.16}
$$
where
$$
P_{12} (\uB\otimes \vB)=\vB\otimes \uB  \tag{2.17}
$$
is the transpostion endomorphism. This defines a Poisson bracket structure on the
space $\Ggl_{rat}(r,\CB)$ of rational $r \times r$ matrix valued functions of 
$\l$.   The $\a_i$'s and $\b_a$'s may be viewed as multi--time parameters
associated to the $n+r$ nonautonomous Hamiltonians $\{H_i, K_a\}_{i=1,\dots
n\atop a=1,\dots r}$ defined by 
$$
\align
H_i &:= {1\over 4 \pi i} \oint_{\l=\a_i} \tr\left( \NN^2(\l)\right)d\l 
=\tr(B N_i) + \sum_{j=1\atop j\neq i}^n{\tr (N_i N_j) \over \a_i -\a_j} 
\tag{2.18a}\\
K_a &:= {1\over 4 \pi i} \oint_{\l=\infty} d\l \l 
\oint_{z=\b_a}dz \left[ \tr\left( (B- z\IB)^{-1} \NN(\l)\right)^2 
-z \tr\left((B-z \IB)^{-1} \NN(\l)\right)\right] \\
&= \sum_{j=1}^n\a_j (N_j)_{aa} + \sum_{n=1 \atop b\neq a}^r 
{(N_\infty)_{ab}(N_\infty)_{ba} \over \b_a - \b_b}. \tag{2.18b}
\endalign
$$
Defining a $1$--form $\th$ on the parameter space by
$$
\th := \sum_{i=1}^n H_i d\a_i  + \sum_{a=1}^r K_a d\b_a,  \tag{2.19}
$$ 
the following results, which may be verified directly [JMU], are a 
consequence of the general $R$--matrix theory, adapted to the nonautonomous
setting [H].  \proclaim
{Theorem} 
 The isomonodromic deformation equations (2.13) are Hamilton's equations
for the set of Hamiltonians $\{H_i, K_a\}$ contained in the $1$--form $\th$
$$
dN_i :=  \sum_{j=1}^n {\di N_i \over \di \a_j}d\a_j +
    \sum_{a=1}^r d\b_a{\di N_i \over \di \b_a} =\{N_i, \ \th\}.  \tag{2.20} 
$$
Furthermore, the Hamiltonians $\{H_i, K_a\}$ all Poisson commute amongst
themselves, and hence $\th$ is a closed form on the parameter space 
$$
d\th =0,  \tag{2.21}
$$
and there exists, at least locally, a function $\t$  (the {\it tau} function)
such that 
$$
\th = d(\ln \t).  \tag{2.22}
$$
\endproclaim

   The Hamiltonian equations determining these isomonodromic deformations may in 
fact  be lifted to an associated symplectic vector space, and then projected
to {\it another} space of rational differential operators, giving rise to a
second, {\it dual} representation of these equations as isomonodromic
deformations  [H].  It is easy to see that all $\NN(\l)$'s of the form (2.10)
may be expressed as 
$$
\NN(\l)= B+ G^T(A- \l \IB)^{-1} F,  \tag{2.23}
$$
where $A$ is a diagonal $N \times N$  matrix, 
$$
A:= \diag(\a_1, \dots, \a_n), \tag{2.24}
$$
with $N=\sum_{i=1}^n k_i$,  $k_i=\text{rk}(N_i)$,  whose eigenvalues  
$\{\a_i\}$  are at the poles of $\NN(\l)$, and have multiplicity $k_i$, and
$(F, G)$ are a  pair of rectangular  $N\times r$ matrices. If the space
$\MM^{N\times r}:=\{(F,G)\}$ of  such pairs is given the canonical symplectic
structure   $$ \o =\tr(dF^T \wedge dG), \tag{2.25}
$$
the map $J^A_B:\MM^{N\times r} \ra \Ggl_{rat}(r,\CB)$ to the space 
$\Ggl_{rat} (r,\CB)$ defined,  for fixed $A$ and $B$, by 
$$
J^A_B: (F,G) \mt \NN(\l),  \tag{2.26}
$$
is a Poisson map. In fact, it may be viewed as defining a Poisson quotient of
$\MM^{N\times r}$ by the canonical Hamiltonian action of the stability 
subgroup $\GGG_A \ss \GGL(N, \CB)$ of $A$, given by 
$$
\align
\GGG_A \times \MM^{N\times r} &\lra \MM^{N \times r}  \\
(g, (F,G)) & \lmt (gF, (g^T)^{-1}G).  \tag{2.27}
\endalign
$$
(The $k_i\times r$ blocks $(F_i, G_i)$ within the matrices $F$ and $G$ 
corresponding to the eigenvalues $\a_i$ determine the local monodromy of the
operator $\DD_\l$ about these points.) Denoting the pull--back of the 
$1$--form $\th$  under $J^A_B$  by $\wt{\th}$, the isomononodromic
deformation equations (2.13) may be viewed as the projection to
$\Ggl_{rat}(r,\CB)$ of the following nonautonomous Hamiltonian equations  on
the auxiliary space $\MM^{N\times r}$:  
$$
dF = \{F, \wt{\th}\}, \quad dG = \{G, \wt{\th}\},  \tag{2.28}
$$
 where
$$
d =  \sum_{j=1}^n d\a_j{\di \over \di \a_j} +
    \sum_{a=1}^r d\b_a {\di  \over \di \b_a}.  \tag{2.29}
$$

  To obtain the {\it dual} set of isomonodromic deformation equations, we 
define a new Hamiltonian quotient of $\MM^{N\times r}$, this time by the
stabilizer  $\GGG_B\ss \GGL(r,\CB)$ of the element $B \in \GGL(r)$, by the
dual Poisson map  $J^B_A:(F,G) \ra \MM(z)$, where 
$$
\MM(z) := A + F(B- z\IB_r)^{-1}G^T = A + \sum_{a=1}^r {M_a\over z- \b_a}. 
\tag{2.30} $$ 
Defining the associated set of differential operators
$$
\align
\wt{\DD}_z &:= {\di \over \di z} -\MM(z) \tag{2.31a}\\
\wt{\DD}_{\a_i}&:= {\di \over \di \a_i} - z E_i - 
\sum_{j=1 \atop j\neq i}^n {E_i M_\infty E_j + E_j M_\infty E_i \over \a_i - \a_j} 
\tag{2.31b}\\
\wt{\DD}_{\b_a} &:= {\di \over \di \b_a} + {M_a \over z- \b_a}, \tag{2.31c}
\endalign
$$
 Hamilton's equations (2.28) are projectible under the map $J^B_A$ and 
imply the commutativity conditions
$$
[\wt{\DD}_{\a_i}, \ \wt{\DD}_{\a_j}]=[\wt{\DD}_{\a_i}, \wt{\DD}_{\b_a}] = 
[\wt{\DD}_{\b_a},\ \wt{\DD}_{\b_b}] =  [\wt{\DD}_z, \wt{\DD}_{\a_i}] = 
[\wt{\DD}_z,\wt{\DD}_{\b_a}] = 0, \tag{2.32} 
$$
which, in turn,  imply the invariance of the monodromy data of the operator
$\wt{\DD}_z$ (resp. $\wt{\DD}_x$) under changes inthe parameter $x$  (resp.
$z$).  These equations may also be interpreted as Hamiltonian equations on
the space of rational $N\times N$ matrix valued functions of $\MM(z)$,
generated by the set of Poisson commuting, nonautonomous Hamiltonians
obtained by projection:  $$
\align
\wt{K}_a &:= {1\over 4 \pi i} \oint_{z=\b_a} \tr\left( \MM^2(z)\right)dz 
=\tr(A B_a) + \sum_{b=1\atop b\neq a}^r{\tr (M_a M_b) \over \b_a -\b_b} 
\tag{2.33a}\\
\wt{H}_i &:= {1\over 4 \pi i} \oint_{z=\infty} dz z 
\oint_{\l=\a_i}d\l \left[ \tr\left( (A- \l \IB_N)^{-1} \MM(z)\right)^2 
-\l \tr\left((A- \l \IB_N)^{-1}\MM(z)\right)\right] \\
&= \sum_{b=1}^r\b_b (M_b)_{ii} + \sum_{j=1 \atop j\neq i}^n 
{(M_\infty)_{ij}(M_\infty)_{ji} \over \a_i - \a_j}. \tag{2.33b}
\endalign
$$
The duality map
$$
(F, G, A, B, \l, z) \ra (G^T, F^T, B, A, z, \l)  \tag{2.34}
$$
may be viewed, after projection, as transforming the system of isomonodromic
deformations equations (2.13) into the dual system (2.32).

   Tthis duality map is reminiscent of the bispectral involution
(1.23) expressed in terms of the matrices $(X,Z)$ determining the 
Calogero--Moser system.  A full explanation of the relationship between these
two maps is not  yet developed, but one certainly exists. What will be
provided in the next section will be a reformulation of the two examples of
bispectral pairs given in section 1 in terms of an equivalent, dual pair of
isomonodromic deformation equations - in the very particular case where the
monodromy happens to be trivial.  These examples in fact illustrate a
correspondence that can be made between  the bispectral pairs of the type
studied by Wilson; i.e., thoe belonging to  rank $1$ bispectral algebras, and
dual pairs of parametric  families of matrix differential operators of first
order, depending rationally on both parameters, and having trivial monodromy.

 \bigskip
\subheading{3. Relation between Bispectrality and Dual Isomonodromy} 
\nopagebreak\smallskip

  The general structure of bispectral equations and infinitesimal 
isomonodromic deformation equations, both involving a wave function depending
on two variables that simultaneously satisfies a linear differential equation
in each of  them, suggests that the two may be related. The similarity
between the duality map (2.34) and the bisectral involution (1.23) further
suggests a relation  between bispectrality and duality. The deeper relation
between bispectral pairs and dual isomonodromic deformations involves the
theory of dressing transformations and the tau function, and will be not
developed here. However, to demonstrate that such a relation does exist, we
consider the two examples of bispectral rank $1$ operators introduced in
section 1.2, and show how an equivalent infinitesimal isomonodromic
deformation system can be derived for each. \medskip
 \line{3.1 \quad{\it Examples.}\hfil}
 \nopagebreak

\noindent  (i) \ 
For the case of the Baker type bispectral wave function 
$$
\psi(x,z) = e^{xz} \left( 1 - {1\over (x+\a)z}\right)  \tag{3.1}
$$
there exists a {\it second} bispectral wave function for the same pair of 
operators $L$ and $\L$ given in (1.24c,d), namely
$$
\psi_1 := \psi(-x-2\a, z) = e^{-2\a z} \psi (x, -z)  \tag{3.2}
$$
Forming the Wronskian matrix
$$
\Psi:= \pmatrix \psi & \psi_1 \\ \psi_x & \psi_{1,x}
        \endpmatrix,  \tag{3.3}
$$
we may associate the following pair of rational $2\times2$ covariant 
derivative operators
$$
\align
\DD_x &:= {\di \over \di x} - 
   \pmatrix 0 & 1 \\ z^2 + {2\over (\a + x)^2} & 0 \endpmatrix  
\tag{3.4a}\\  
\DD_z &:= {\di \over \di z} -  \pmatrix -\a & {\a + x \over z}  \\ 
(\a+x)z + {2\over (\a+x)z} & {1\over z} -\a \endpmatrix,   \tag{3.4b}
\endalign
$$
which simultaneously annihilate $\Psi$,
$$
\DD_x \Psi=0, \qquad \DD_z \Psi =0,  \tag{3.5}
$$
This is equivalent to the bispectral conditions (1.1a,b) for the operators 
$L,\ \L$ defined in (1.24c,d) and implies the commutativity condition
$$
[\DD_x, \ \DD_z]=0.  \tag{3.6}
$$
It follows that the monodromy of either of these operators, say $\DD_z$, is 
invariant under the deformations determined by varying the other parameter, 
$x$.  Note however the fact that there is not just {\it one} single--valued
bispectral wave function, but a {\it basis} of single valued bispectral wave
functions,  which  implies in this case that the operators $\DD_x$ and
$\DD_z$ actually have  {\it trivial} monodromy for all parameter values.

  To obtain the {\it dual} system of (trivial) isomonodromic deformation 
equations, we just define new operators similarly, using the Wronskian matrix
formed by differentiating with respect to the variable $z$:
$$
\wt{\Psi}:= \pmatrix \psi & \psi_1 \\ \psi_z & \psi_{1,z}
        \endpmatrix.  \tag{3.7}
$$
This is simultaneously annihilated by the operators
$$
\align
\wt{\DD}_x &:= {\di \over \di x} - 
   \pmatrix {\a z\over \a +x} & {z\over \a +x} \\ 
{(x^2 +2\a x)z\over \a+x}+ {2+\a z\over (\a +x)z} & {1-\a z\over \a + x}
\endpmatrix    \tag{3.8a}\\  
\wt{\DD}_z &:= {\di \over \di z} - 
   \pmatrix   0 &  1 \\ 
{2\over z^2} + x^2 + 2\a x & -2\a \endpmatrix,   \tag{3.8b}
\endalign
$$
implying the commutativity condition
$$
[\wt{\DD}_x, \ \wt{\DD}_z]=0,  \tag{3.9}
$$
from which again follows that the monodromy of either of these operators
is invariant under the deformations determined by varying the other parameter.
 
Note however that the singularity structure of the operators $\DD_x, \DD_z,
\wt{\DD_x}$ and $\wt{\DD}_z$ is such that the Hamiltonian $R$--matrix
formulation discussed in section 2.2 does not suffice. In order to include
these, the  $R$--matrix approach must be extended (cf.  [HTW], [HR], [HW1],
[HW2]) to allow for singularities of higher order at $\infty$. 

\noindent  (ii) \
For the case of the bispectral wave function
$$ 
\psi_{W(X,Z)}(x,z) = e^{xz} \left(1 - {2\over x z} + {2\over x^2 z^2}\right),
\tag{3.10}
$$
because of the invariance of the eigenvector equations for the operators
$L$ and $\L$  in (1.25c,d) under the cyclic group generated by
$$
z\mt \o z, \quad \o:= e^{2\pi i\over 3},  \tag{3.11}
$$
we may define a basis $\{\psi_0, \psi_1, \psi_2\}$ of bispectral wave
functions by $$
\psi_j(x,z) := \psi(x, \o^j z), \quad j=0,1,2. \tag{3.12}
$$
The Wronskian matrix
$$
\Psi:= \pmatrix \psi_0 & \psi_1 & \psi_2 \\
    \psi_{0,x} & \psi_{1,x}  & \psi_{2,x} \\
   \psi_{0,xx} & \psi_{1,xx}  & \psi_{2,xx} 
        \endpmatrix  \tag{3.13}
$$
is then simultaneously annihilated by the operators
$$
\align
\DD_x &:= {\di \over \di x} - 
   \pmatrix 0 & 1 & 0 \\ 
  0  &  0  &  1  \\
  z^3  - {12\over x^3} & {6\over x^2} &  0  
 \endpmatrix    \tag{3.14a}
\\ 
\DD_z &:= {\di \over \di z} - 
   \pmatrix 0 & {x \over z} & 0 \\ 
0 & {1\over z}  & {x\over z} \\
x z^2 - {12\over x^2 z}  & {6\over xz}  & {2\over z}
\endpmatrix.   \tag{3.14b}
\endalign
$$
This is equivalent to the eigenvector equations for the operators
$L$ and $\L$ of eqs. (1.25c,d), and implies the commutativity of the
operators  $\DD_x, \DD_z$. Hence the monodromy of each of these operators
(which again is trivial because of the single valuedness of the matrix
$\Psi$), remains invariant under changes in the parameter values. Because of
the symmetry of the bispectral wave functions under the interchange
$x\leftrightarrow z$, the corresponding ``dual'' isomonodromic system is
obtained by simply interchanging the two variables in the definitions of the
operators  $\DD_x, \ \DD_z$. As in the previous example, the proper treatment
of these systems within the $R$--matrix framework requires the inclusion of
higher order singularities at $\infty$.  

   Proceeding similarly, we may associate to every rank $1$ bispectral pair
two mutually dual $1$--parameter families of differential operators having
trivial  monodromy for all parameter values, since in  each case we may form
a basis of single--valued bispectral wave functions. This procedure may also
be applied to  the case of higher rank bispectral operators, except that the
monodromy of the resulting dual families of operators is no longer
necessarily trivial, and we obtain instead $1$-parameter families of
operators with constant monodromy under variation of the deformation
parameters. These more general cases, as well as their relations to the
evaluation of certain limits of Fredholm determinants, will be dealt with
elsewhere.

 \bigskip \bigskip 
%%%%%%%%%%%%%%%%%%%%%%%%%%%%%%% Acknowledgements %%%%%%%%%%%%%%%%%%%%%%%%%%
\noindent{\it Acknowledgements.} The author would like to thank
A. Gr\"unbaum, A. Kasman and A. Orlov for helpful discussions. 
\bigskip\bigskip  
%%%%%%%%%%%%%%%%%%%%%%%%%%%%%%%% References %%%%%%%%%%%%%%%%%%%%%%%%%%%%%%%
\centerline{\bf References} \nobreak \bigskip  {\smaller
\item{\bf [AMM]} H. Airault, H.P. McKean,and J. Moser, ``Rational and
Elliptic Solutions od the Korteweg--de Vries Equation and a Related
Many--Body Problem, {\it Commun. Pure Appl. Math.} {\bf 30}, 95--148 (1977).
\item{\bf [DG]} J.J. Duistermaat and A. Gr\"unbaum, ``Differential equations
in the Spectral parameter'', {\it Commun. Math.  Phys.}, {\bf 107}, 177--2404
(1986). %
\item{\bf [DJKM]}  E. Date, M. Jimbo, M. Kashiwara, M. and T. Miwa, 
``Transformations Groups for Soliton Equations''  {\it I. Proc. Japan. Acad. }
{\bf 57A}, 342--392 (1981); {\it II. Ibid.} 387--392 (1981); {\it III. J.
Phys. Soc. Japan} {\bf 50}, 3806--3812 (1981); {\it IV. Physica} {\bf 4D},
343--365 (1982); {\it Publ. RIMS, Kyoto Univ.} {\bf 18}, 1111--1119 (1982);
{\it VI. J. Phys. Soc. Japan} {\bf 50}, 3813--3818 (1981); {\it Publ. RIMS,
Kyoto Univ.}{\bf 18}, 1077--1110 (1982). %
\item{\bf [H]} J. Harnad, ``Dual Isomonodromic Deformations and Moment Maps 
to Loop Algebras'', {\it Commun. Math. Phys.} {\bf 166}, 337--365 (1994).
\item {\bf [HR]}	J. Harnad and M. Routhier ,  ``R--matrix Construction of
Electromagnetic Models  for the Painlev\'e Transcendents'', {\it J. Math.
Phys.}  {\bf 36}, 4863--4881 (1995). 
\item{\bf [HTW]}	J. Harnad,  C.A. Tracy,  and H. Widom, 
``Hamiltonian Structure of Equations Appearing in 	Random Matrices'' , in:
{\it  Low Dimensional Topology and Quantum Field Theory}, ed. H. Osborn, 
(Plenum, New York  1993), pp. 231--245.
\item{\bf [HW1]}	J. Harnad and M.--A. Wisse, ``Loop Algebra Moment Maps and 
Hamiltonian Models for the	Painlev\'e Transcendants'', in: Mechanics Day
Workshop Proceedings, ed. P.S. Krishnaprasad, T. Ratiu and W.F. Shadwick,  
{\it AMS--Fields Inst. Commun.} {\bf 7}, 155--169 (1996).  
\item{\bf [HW2]}	J. Harnad and M.--A. Wisse,
``Moment Maps to Loop Algebras, Classical $R$--Matrix and Integrable 
	Systems'', in: {\it Quantum Groups, Integrable Models and Statistical Systems}, 
Proc. of the 1992  CAP 	Summer Workshop , ed. L. Vinet and J. Letourneux 
(World Scientific, Singapore, 1993). 
 \item{\bf[JMMS]} M. Jimbo, T. Miwa, Y.  M\^ori and M. Sato, ``Density
Matrix of an Impenetrable Bose Gas and the Fifth Painlev\'e Transcendent'',
{\it Physica} {\bf 1D}, 80-158 (1980). 
 \item{\bf[JMU]} M. Jimbo, T. Miwa, K. Ueno, K., ``Monodromy Preserving
Deformation of Linear Ordinary Differential Equations with Rational
Coeefficients I.'', {\it Physica} {\bf 2D}, 306-352 (1981).  
 \item{\bf[JM]} M. Jimbo and T. Miwa, ``Monodromy Preserving
Deformation of Linear Ordinary Differential Equations with Rational
Coeefficients II, III.'', {\it Physica} {\bf 2D}, 407-448 (1981); {\it ibid.},
{\bf 4D}, 26-46 (1981).
\item{\bf [K]} A. Kasman, ``Bispectral KP Soultions and Linearization of
Calogero--Moser Particle Systems'', {\it Commun. Math. Phys.} {\bf 172},
427--448 (1995).
\item{\bf [Kr]} I.M. Krichever,`` Rational Solutions of the
Kadomtsev--Petviashvili
 Equation and Integrable Systems of $N$ Particles on a Line'',
 {\it Func. Anal. and Appl.} {\bf 12}, 59--61 1978
\item{\bf [M]} J. Moser,  ``Three Integrable Hamiltonian Systems Connected
with Isospectral Deformations'', {\it Adv. Math.} {\bf 16}, 197--220 (1975).
\item{\bf [SS]} M. Sato and Y. Sato, ``Soliton Equations as Dynamical Systems
   on Infinite Dimensional Grassmann Manifold'', 
{\it  Lecture Notes in Num. Appl. Anal.} {\bf  5}, 259-271 (1982).
\item{\bf [SW]} G. Segal and  G. Wilson, 
``Loop Groups and Equations of KdV Type'', {\it Publ. Math. I.H.E.S.}, {\bf
61}, 5--65 (1985). 
\item{\bf [W1]} G. Wilson,
``Bispectral Commutative Ordinary Differential Operators'', {\it J. Reine
Angew. Math.}, {\bf 442}, 179--204 (1993).
\item{\bf [W2]} G. Wilson,
``Collisions of Calogero--Moser Particles and an Adelic Grassmannian'',
preprint (Imperial College, 1996).

}\vfill \eject
\enddocument